\documentclass[final,3p,times]{elsarticle}
\usepackage[T1]{fontenc}       
\usepackage{newtxtext,newtxmath}
\usepackage[utf8]{inputenc}

\usepackage{amssymb}
\usepackage{amsmath}
\usepackage{multicol}
\usepackage[english]{babel}
\usepackage{color}
\usepackage{bm}
\usepackage{natbib}
\usepackage{caption}

\usepackage{booktabs}

\journal{Physica A}

\begin{document}

\begin{frontmatter}

\title{Impact of Lipid Structural Variations on Bilayer Properties: A Coarse-Grained Molecular Dynamics Study}

\author[label1]{Sonam Phuntsho}

\address[label1]{Department of Physical Science, Sherubtse College, Royal University of Bhutan,
            42007 Kanglung,
            Trashigang,
            Bhutan}

\begin{abstract}

The supramolecular assembly of lipids into bilayer membranes is essential for cellular structure and function. However, the impact of lipid structural variations such as acyl chain length, degree of unsaturation, and headgroup type on bilayer properties remains incompletely understood. This study employs coarse-grained molecular dynamics simulations using the Martini force field to investigate seven distinct lipid species, aiming to compute critical bilayer parameters including area per lipid, bilayer thickness, and lateral diffusion coefficients. Our simulations reveal that lipids with longer acyl chains exhibit increased bilayer thickness, while unsaturation introduces kinks in the acyl chains, generally reducing bilayer thickness and increasing the area per lipid. Lipids with unsaturated chains demonstrate higher lateral diffusion coefficients, enhancing membrane fluidity. Variations in headgroup chemistry significantly influence lipid packing and membrane dynamics. This investigation advances our understanding of membrane biophysics and has significant implications for the design of lipid-based systems in biomedical applications.

\end{abstract}


\end{frontmatter}

\section{Introduction}
The supramolecular assembly of lipids into bilayer membranes is fundamental to the structure and function of biological cells \cite{Alberts2014}. Lipid bilayers not only provide structural integrity to cell membranes but also play critical roles in processes such as signaling, transport, and energy transduction \cite{Singer1972, M, R}. Understanding the biophysical properties of lipid bilayers including area per lipid, bilayer thickness, and lateral diffusion coefficients is essential for elucidating membrane functionality and interactions with proteins and other biomolecules \cite{Nagle2000, Ma, H, G}.

Molecular dynamics (MD) simulations have become indispensable tools for studying lipid bilayers at the molecular level \cite{Tieleman2006}. However, all-atom MD simulations are computationally intensive, especially for large systems or long timescales \cite{Marrink2013}. Coarse-grained (CG) MD simulations offer a viable alternative by simplifying the system while retaining essential physical characteristics \cite{Ingolfsson2014}. The Martini force field is one of the most widely used CG models for biomolecular simulations, effectively balancing computational efficiency and accuracy \cite{Marrink2007, Monticelli2008, Periole2009}.
Despite advances in MD simulations, most studies have focused on standard phospholipids like dipalmitoylphosphatidylcholine (DPPC) and dioleoylphosphatidylethanolamine (DOPE) \cite{deVries2005,Marrink2004}. There is limited research on lipids with unusual chain lengths, degrees of unsaturation, or different headgroups \cite{Scott2002,Khandelia2009}. The influence of lipid chain length and saturation significantly affects bilayer thickness and fluidity, with longer chains and higher saturation levels generally leading to thicker and less fluid bilayers \cite{Schäfer2011, Hofsaess2003,Bacle2017}. However, systematic investigations comparing a range of lipids differing in chain length, saturation, and headgroup are limited \cite{Pan2009, Kucerka2011}, and the impact on lateral diffusion coefficients---a key parameter affecting membrane dynamics---remains underexplored \cite{Kucerka2005, Filippov2003}.

This study aims to compute the area per lipid, bilayer thickness, and lateral diffusion coefficients for bilayers composed of eight different lipids using coarse-grained MD simulations with the Martini force field \cite{Marrink2007}. We analyze the effects of variations in acyl chain length, degree of unsaturation, and headgroup type on these bilayer properties. The simulation results are compared with available experimental data and previous simulation studies to validate the findings.
Understanding how lipid structural variations influence bilayer properties is crucial for enhancing our fundamental knowledge of membrane biophysics and aiding in the interpretation of experimental data \cite{Simons1997, Carpenter2014, Sezgin2017}. Such insights are valuable for the design of lipid-based drug delivery systems, artificial membranes, and in understanding membrane-associated processes in health and disease \cite{Torchilin2005,Funari2003}. Furthermore, the findings can contribute to refining coarse-grained models like Martini, improving their predictive power for complex lipid systems \cite{deJong2013}.

\section{Methodology}
All molecular dynamics (MD) simulations were performed using the GROMACS 2018.3 software package \cite{VanDerSpoel2005}, employing the Martini coarse-grained (CG) force field version 2.2 \cite{Marrink2007}. A simulation box containing a random configuration of 128 lipid molecules was constructed for each lipid species. The lipids studied included DNPC, DPPC, DTPE, DVPE, DXPE, LPPC, and POPE, modeled using the Martini CG lipid representation, which has been validated for studying lipid bilayer systems \cite{Marrink2004}.

The lipid assemblies were solvated with coarse-grained water molecules at a ratio of six water beads per lipid molecule, corresponding to 24 all-atom waters per lipid and totaling 768 water beads in the system. This solvation level ensures adequate hydration of the lipid bilayer and is consistent with previous Martini simulations \cite{Yesylevskyy2010}.
Equilibration of the system was carried out using a time step of 20\,fs, as recommended for Martini simulations to maintain numerical stability while capturing essential dynamics \cite{Marrink2007}. The total simulation time was set to 20\,ns (1,000,000 steps). The neighbor list was updated every 10 steps using the Verlet cutoff scheme with a buffer tolerance of 0.005\,kJ\,mol$^{-1}$\,ps$^{-1}$, optimizing computational efficiency without sacrificing accuracy \cite{Pall2013, Neale2016}.

Non-bonded interactions were calculated using a cutoff of 1.1\,nm for both Coulombic and van der Waals interactions. The reaction-field method was employed for electrostatics with a relative dielectric constant ($\epsilon_r$) of 15, aligning with Martini force field recommendations \cite{Marrink2007}. This approach effectively accounts for the screening effect in a polarizable medium without introducing long-range electrostatic artifacts \cite{Patra2003}.

\begin{minipage}{0.4\textwidth}
    \centering
    \begin{equation*}
V_{\mathrm{LJ}}(r_{ij}) = 4\epsilon_{ij} \left[ \left(\frac{\sigma_{ij}}{r_{ij}}\right)^{12} - \left(\frac{\sigma_{ij}}{r_{ij}}\right)^{6} \right]
\end{equation*}
\end{minipage}
\begin{minipage}{0.5\textwidth}
\centering
\begin{equation*}
V_{\text{RF}}(r_{ij}) = 
\begin{cases}
\displaystyle \frac{q_i q_j}{4 \pi \epsilon_0}\left[\frac{1}{r_{ij}} + \frac{\epsilon_r - 1}{2\epsilon_r + 1}\frac{r_{ij}^{2}}{r_c^3}\right], & r_{ij} < r_c \\[8pt]
0, & r_{ij} \ge r_c
\end{cases}
\end{equation*}
\end{minipage}
\\ \\ 
Temperature coupling was implemented using the velocity-rescale thermostat \cite{Bussi2007} with a coupling time constant ($\tau_t$) of 1.0\,ps for both lipid and water groups, maintaining the system at 323\,K. Pressure coupling was applied semi-isotropically using the Parrinello–Rahman barostat \cite{Parrinello1981}, suitable for anisotropic systems like lipid bilayers, with a coupling constant ($\tau_p$) of 12.0\,ps and a reference pressure of 1.0\,bar. A compressibility of $3 \times 10^{-4}$\,bar$^{-1}$ was used, appropriate for simulating biological membranes \cite{Patra2003}.
Initial velocities were generated according to a Maxwell–Boltzmann distribution at 323\,K with a random seed to ensure reproducibility. Bond constraints were applied using the LINCS algorithm \cite{Hess1997} to maintain bond geometries during the simulation.


\section{Results and Discussion}

\subsection{Area per Lipid}
The area per lipid is a fundamental parameter influencing the physical properties of lipid bilayers, including membrane fluidity, thickness, and permeability \cite{Nagle2000}. Using coarse-grained molecular dynamics simulations, we calculated the area per lipid for eight different lipid species. The results are summarized in Table \ref{table:area} below.

\begin{table}[htbp]
    \centering
    \begin{minipage}{0.4\textwidth}
    \caption{Calculated area per lipid for various lipid species.}
    \label{table:area}
    \begin{tabular}{lc}
        \toprule
        \textbf{Lipid} & \textbf{Area per Lipid (nm$^2$)} \\
        \midrule
        DNPC & 0.74 \\
        DPPC & 0.63 \\
        DTPE & 0.59 \\
        DVPE & 0.66 \\
        DXPE & 0.68 \\
        LPPC & 0.62 \\
        POPE & 0.64 \\
        \bottomrule
    \end{tabular}
    \end{minipage}
    \hfill
    \begin{minipage}{0.55\textwidth}
    \centering
	\includegraphics[width=\linewidth]{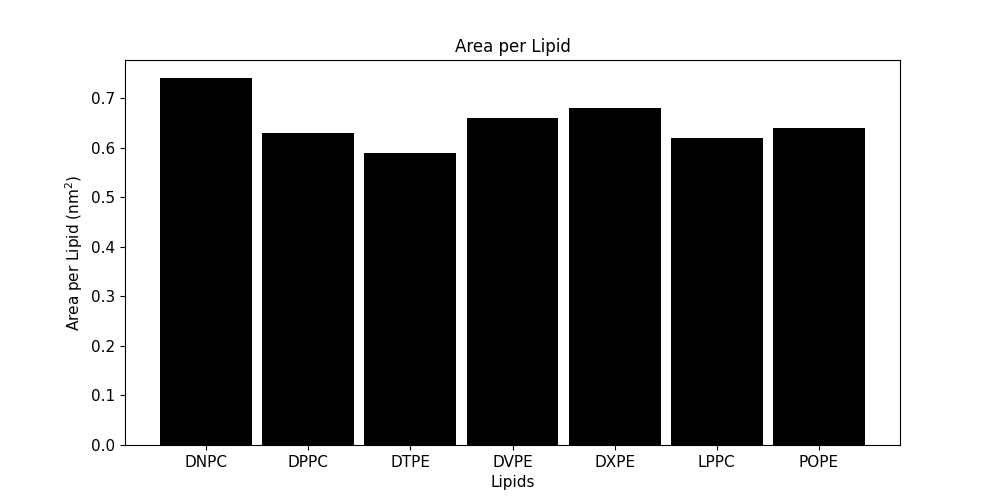}
    \end{minipage}
\end{table}

Our simulations indicate that DNPC exhibits the largest area per lipid (0.74\,nm$^2$) compared to other lipids. This variation can be attributed to differences in acyl chain length and degree of unsaturation. DNPC contains very long acyl chains, which increase the area per lipid due to the larger space occupied by the extended chains \cite{Hofsaess2003}. Additionally, unsaturated chains introduce kinks in the hydrocarbon chains due to cis double bonds, leading to increased area per lipid because the kinks prevent tight packing \cite{Hofsaess2003}.

The influence of headgroup type is evident when comparing lipids like DPPC and POPE. DPPC, a phosphatidylcholine (PC), has an area per lipid of 0.63\,nm$^2$, while POPE, a phosphatidylethanolamine (PE), has an area of 0.64\,nm$^2$. PE headgroups are smaller and can form stronger intermolecular hydrogen bonds, leading to tighter packing \cite{Lewis1983}. However, the unsaturation in POPE's acyl chains can offset this effect, resulting in a similar or slightly increased area per lipid compared to DPPC.

Chain length also plays a significant role. For instance, DTPE (short saturated chains) has an area per lipid of 0.59\,nm$^2$, whereas DXPE (long saturated chains) shows a larger area of 0.68\,nm$^2$. Longer chains increase van der Waals interactions among lipids, which can expand the bilayer and increase the area per lipid \cite{Bacle2017}. This observation aligns with previous studies indicating that membrane thickness and area per lipid increase with acyl chain length \cite{Pan2009}.
The calculated area per lipid for DPPC (0.63\,nm$^2$) is in reasonable agreement with experimental values ranging from 0.62 to 0.64\,nm$^2$ \cite{Nagle2000,Kucerka2011}. 


\subsection{Lateral Diffusion Coefficients}
The lateral diffusion coefficients of the lipid species were determined from mean squared displacement (MSD) analysis. The calculated diffusion coefficients are presented in Table \ref{table:diffusion} below. These lipids differ in acyl chain length, degree of saturation, and headgroup type, factors known to influence membrane dynamics and organization.

\begin{table}[htbp]
    \centering
    \begin{minipage}{0.3\columnwidth}
    \caption{Calculated lateral diffusion coefficients for various lipid species.}
    \label{table:diffusion}
    \begin{tabular}{lc}
        \toprule
        \textbf{Lipid} & \textbf{D ($\times10^{-7}$\,cm$^2$/s)} \\
        \midrule
        DNPC & 4.17 \\
        DPPC & 7.40 \\
        DTPE & 11.99 \\
        DVPE & 8.68 \\
        DXPE & 3.73 \\
        LPPC & 5.71 \\
        POPE & 4.36 \\
        \bottomrule
    \end{tabular}
    \end{minipage}
  \hfill
  \begin{minipage}{0.65\textwidth}
    \centering
	\includegraphics[width=\linewidth]{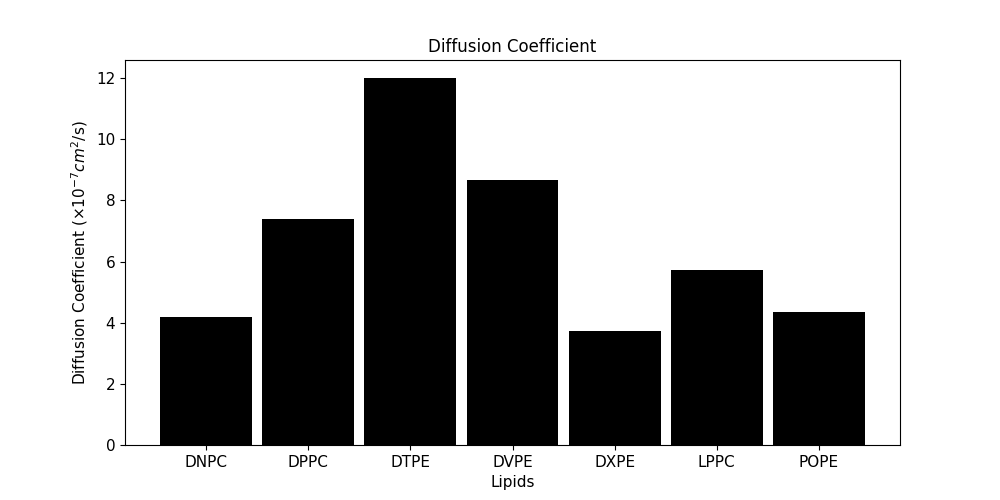}
  \end{minipage}
\end{table}

The diffusion coefficients reveal a clear inverse relationship between acyl chain length and lipid mobility. DTPE, possessing the shortest acyl chains, exhibited the highest diffusion coefficient ($11.99 \times 10^{-7}$ cm$^{2}$/s). The reduced chain length decreases van der Waals interactions between adjacent lipids, resulting in increased membrane fluidity and enhanced lateral diffusion. This observation aligns with previous studies indicating that shorter acyl chains facilitate higher lipid mobility due to decreased hydrophobic interactions within the bilayer core.

Conversely, DXPE, containing the longest saturated acyl chains, showed the lowest diffusion coefficient ($3.73 \times 10^{-7}$ cm$^{2}$/s). The extended chain length enhances van der Waals forces and chain entanglement, leading to a more ordered and rigid membrane structure that hinders lipid movement. This finding is consistent with the established understanding that longer acyl chains contribute to decreased membrane fluidity and lower diffusion rates.

The degree of unsaturation in the acyl chains also significantly affects lipid diffusion. DVPE, with monounsaturated acyl chains (C16:1 and C18:1), demonstrated a higher diffusion coefficient ($8.68 \times 10^{-7}$ cm$^{2}$/s) compared to its saturated counterpart DPPC ($7.40 \times 10^{-7}$ cm$^{2}$/s), which has similar chain lengths. The presence of double bonds introduces kinks in the acyl chains, disrupting the orderly packing of lipids and increasing membrane fluidity. This structural disorder reduces the energy barrier for lateral movement, thereby enhancing diffusion.

Similarly, POPE, which contains one saturated and one monounsaturated acyl chain, exhibited a moderate diffusion coefficient ($4.36 \times 10^{-7}$ cm$^{2}$/s). The combination of saturated and unsaturated chains results in intermediate packing efficiency and fluidity, reflecting in its diffusion behavior.

The headgroup chemistry plays a pivotal role in lipid interactions and membrane properties. Phosphatidylethanolamine (PE) lipids generally promote tighter packing due to their smaller headgroup size compared to phosphatidylcholine (PC) lipids \cite{Lewis1983}. Despite this, DTPE (a PE lipid) showed the highest diffusion coefficient. This suggests that the effect of short acyl chain length in DTPE overrides the headgroup influence, leading to increased diffusion.

LPPC, a PC lipid with asymmetric acyl chains, showed a diffusion coefficient of $5.71 \times 10^{-7}$ cm$^{2}$/s. The presence of a shorter acyl chain increases membrane fluidity, but the larger PC headgroup may contribute to reduced mobility compared to DTPE. DNPC, a PC lipid with very long monounsaturated acyl chains, had a lower diffusion coefficient ($4.17 \times 10^{-7}$ cm$^{2}$/s), highlighting that chain length can have a more pronounced effect than headgroup type in certain contexts \cite{Kucerka2005, Simons1997}.

\subsection{Bilayer Thickness}
\begin{table}[htbp]
    \centering
    \begin{minipage}{0.35\textwidth}
    \caption{Calculated bilayer thicknesses for various lipid species.}
    \label{table:thickness}
    \begin{tabular}{lc}
        \toprule
        \textbf{Lipid} & \textbf{Bilayer Thickness (nm)} \\
        \midrule
        DNPC & 3.80 \\
        DPPC & 3.18 \\
        DTPE & 2.92 \\
        DVPE & 3.37 \\
        DXPE & 3.79 \\
        LPPC & 2.16 \\
        POPE & 3.80 \\
        \bottomrule
    \end{tabular}
     
    \end{minipage}
    \hfill
    \begin{minipage}{0.6\textwidth}
    \centering
	\includegraphics[width=\linewidth]{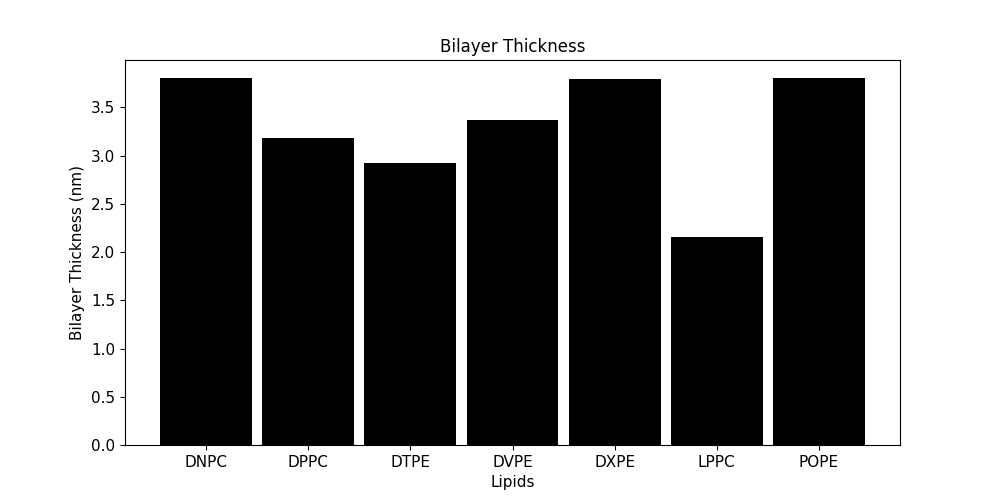}
    \end{minipage}
    
\end{table}

The bilayer thicknesses were determined by calculating the distance between the phosphate groups of the opposing leaflets. The results are summarized in Table \ref{table:thickness} below.

The bilayer thicknesses obtained displayed a clear dependence on acyl chain length. DPPC, consisting of two saturated palmitoyl chains (C16:0), exhibited a bilayer thickness of 3.18\,nm, consistent with previous studies \cite{Alberts2014}. DNPC and POPE, with longer acyl chains, showed increased bilayer thicknesses of 3.80\,nm. The longer acyl chains contribute to increased bilayer thickness due to the extended length of the hydrocarbon tails \cite{Hofsaess2003}.

The effect of unsaturation on bilayer thickness is more complex. Unsaturated chains introduce kinks that can reduce the effective chain length, potentially decreasing bilayer thickness \cite{Hofsaess2003}. However, in lipids with very long chains, such as DNPC, the effect of chain length may dominate over the effect of unsaturation, resulting in an overall increase in bilayer thickness.

The significantly lower bilayer thickness observed for LPPC (2.16\,nm) is notable. LPPC is a lysophospholipid with only one acyl chain, which explains the reduced bilayer thickness. Lysophospholipids are known to form less stable bilayers due to their conical shape and tendency to form micelles rather than bilayers \cite{Lewis1983, Cullis1979}. This could also affect the reliability of the bilayer properties measured for LPPC in our simulations.

\section{Conclusion}
This study employed coarse-grained molecular dynamics simulations with the Martini force field to investigate how variations in acyl chain length, degree of unsaturation, and headgroup type affect the supramolecular assembly and properties of lipid bilayers. By analyzing eight different lipid species---DNPC, DPPC, DTPE, DVPE, DXPE, LPPC, and POPE---we computed key bilayer characteristics such as area per lipid, bilayer thickness, and lateral diffusion coefficients.

Our findings reveal that acyl chain length significantly influences bilayer properties. Lipids with longer acyl chains, like DNPC and POPE, exhibited increased bilayer thicknesses due to the extended length of their hydrocarbon tails. Unsaturation introduced kinks in the acyl chains, leading to increased area per lipid and enhanced membrane fluidity, as evidenced by higher lateral diffusion coefficients. However, the effect of unsaturation on bilayer thickness was less straightforward; while kinks can reduce effective chain length, in lipids with very long chains, the chain length effect may dominate, resulting in increased bilayer thickness.


The headgroup chemistry played a crucial role; phosphatidylethanolamine (PE) headgroups formed stronger intermolecular hydrogen bonds compared to phosphatidylcholine (PC) headgroups, leading to tighter packing and affecting both area per lipid and diffusion properties.

Comparisons with experimental data and previous simulations validated our results, confirming the reliability of the Martini coarse-grained force field in capturing essential bilayer properties. For instance, the calculated area per lipid for DPPC and POPE closely matched experimental values, underscoring the model's accuracy.


The insights gained enhance our fundamental knowledge of membrane biophysics, aiding in the interpretation of experimental data related to membrane dynamics and structure. These findings have practical implications for the design of lipid-based drug delivery systems and artificial membranes, as well as for understanding membrane-associated processes in health and disease.
Furthermore, the study contributes to refining coarse-grained models like Martini, improving their predictive power for complex lipid systems. By demonstrating the model's effectiveness across a diverse set of lipids, we pave the way for future simulations involving even more complex membrane compositions.
In summary, variations in acyl chain length, unsaturation, and headgroup chemistry profoundly modulate bilayer properties such as area per lipid, bilayer thickness, and lateral diffusion coefficients. These findings are instrumental in advancing our comprehension of membrane structure-function relationships and hold significant relevance for both biological research and biotechnological applications.



\begin{thebibliography}{99}
    \bibitem{Alberts2014} Alberts, B. \textit{Molecular Biology of the Cell}. \textit{Garland Science}. (2014).
    \bibitem{Singer1972} Singer, S. J., \& Nicolson, G. L. The fluid mosaic model of the structure of cell membranes. \textit{Science}. \textbf{175}, 720-731. (1972).
    
    \bibitem{M} Mueller, P. \& Rudin, D. O. Translocators in bimolecular lipid membranes: their role in dissipative and conservative bioenergy transductions. \textit{Current Topics in Bioenergetics}, \textbf{3}, 157-249, (1969).
    
    \bibitem{R} Risselada, H. J., \& Marrink, S. J. The molecular face of lipid rafts in model membranes. \textit{Proc. Natl. Acad. Sci.} \textbf{105}, 17367–17372, (2008).

    \bibitem{Nagle2000} Nagle, J. F., \& Tristram-Nagle, S. Structure of lipid bilayers. \textit{Biochimica et Biophysica Acta (BBA) - Reviews on Biomembranes}. \textbf{1469}, 159-195, (2000).
    \bibitem{Ma} Marrink, S. J., \& Mark, A. E. The mechanism of vesicle fusion as revealed by molecular dynamics simulations. \textit{J. Am. Chem. Soc.}, \textbf{125}, 11144–11145, (2003).
    
    \bibitem{H} Hakobyan, D. \& Heuer, A. (2014). Key molecular requirements for raft formation in lipid/cholesterol membranes. \textit{PLoS ONE}. \textbf{9}, e87369, (2014).
    
    \bibitem{G} Helgi I. I. \& Andersen, O. S. Alcohol's Effects on Lipid Bilayer Properties. \textit{Biophysical Journal}. \textbf{101}, 847-855, (2011).
    
    \bibitem{Tieleman2006} Tieleman, D. P., \& Marrink, S. J. (2006). Lipids out of equilibrium: energetics of desorption and pore mediated flip-flop. \textit{J. Am. Chem. Soc.} \textbf{128}, 12462-12467, (2006).
    \bibitem{Marrink2013} Marrink, S. J., \& Tieleman, D. P. Perspective on the Martini model. \textit{Chem. Soc. Rev.} \textbf{42}, 6801-6822, (2013).
    \bibitem{Ingolfsson2014} Ingólfsson, H. I., Lopez, C. A., Uusitalo, J. J., de Jong, D. H., Gopal, S. M., Periole, X., \& Marrink, S. J. The power of coarse graining in biomolecular simulations. \textit{Wiley Interdisciplinary Reviews: Computational Molecular Science}. \textbf{4}, 225-248, (2014)
    
    \bibitem{Marrink2007} Marrink, S. J., Risselada, H. J., Yefimov, S., Tieleman, D. P., \& de Vries, A. H. (2007). The MARTINI force field: Coarse grained model for biomolecular simulations. \textit{J. Phys. Chem. B}, \textbf{111}, 7812-7824, (2007).
    
    \bibitem{Monticelli2008} Monticelli, L., Kandasamy, S. K., Periole, X., Larson, R. G., Tieleman, D. P., \& Marrink, S. J. The MARTINI coarse-grained force field: Extension to proteins. \textit{ J. Chem. Theory and Comput.} \textbf{4}, 819-834, (2008).
    
    \bibitem{Periole2009} Periole, X., Cavalli, M., Marrink, S. J., \& Ceruso, M. A. Combining an elastic network with a coarse-grained molecular force field: Structure, dynamics, and intermolecular recognition. \textit{ J. Chem. Theory Comput.} \textbf{5}, 2531-2543, (2009)
    
    \bibitem{deVries2005} Gumbart, J., Wang, Y., Aksimentiev, A., Tajkhorshid, E. \& Schulten, K. Molecular dynamics simulations of proteins in lipid bilayers. \textit{Current Opinion in Structural Biology}. \textbf{15}, 423-431, (2005).
    
    \bibitem{Marrink2004} Marrink, S. J., de Vries, A. H., \& Mark, A. E. (2004). Coarse grained model for semiquantitative lipid simulations. \textit{J. Phys. Chem. B}. \textbf{108}, 750-760, (2004).
    
    \bibitem{Scott2002} Scott, H. L. Modeling the lipid component of membranes. \textit{Current Opinion in Structural Biology}. \textbf{12}, 495-502, (2002).
    
    \bibitem{Khandelia2009} Khandelia, H., \& Mouritsen, O. G. Lipid gymnastics: Evidence of complete acyl chain reversal in oxidized phospholipids from molecular simulations. \textit{Biophysical Journal} \textbf{96}, 2734-2743, (2009).
    
    \bibitem{Schäfer2011} Schäfer, L. V., de Jong, D. H., Holt, A., Rzepiela, A. J., de Vries, A. H., Poolman, B., Killian, J. A., \& Marrink, S. J. Lipid packing drives the segregation of transmembrane helices into disordered lipid domains in model membranes. \textit{Proc. Natl. Acad. Sci.} \textbf{108}, 1343-1348, (2011).
    
    \bibitem{Hofsaess2003} Hofsäss, C., Lindahl, E., \& Edholm, O. Molecular dynamics simulations of phospholipid bilayers with cholesterol. \textit{Biophysical Journal}. \textbf{84}, 2192-2206, (2003).
    
    \bibitem{Bacle2017} Baccouch, R., Shi, Y.,  Vernay, E., Mathelié-Guinlet, M., Taib-Maamar, N., Villette, S., Feuillie, C., Rascol, E., Nuss, P., Lecomte, S., Molinari, M., Staneva, G. \& D. Alves, I. D., (2017).  The impact of lipid polyunsaturation on the physical and mechanical properties of lipid membranes. \textit{Biochimica et Biophysica Acta (BBA) - Biomembranes}. \textbf{1865}, 184084, (2023).
    
    \bibitem{Pan2009} Pan, J., Tristram-Nagle, S. \& Nagle, J. F. (2009). Effect of cholesterol on structural and mechanical properties of membranes depends on lipid chain saturation. \textit{Phys. Rev. E}. \textbf{80}, 021931, (2009).
    
    \bibitem{Kucerka2011} Kucerka, N., Nieh, M. P., \& Katsaras, J. Fluid phase lipid areas and bilayer thicknesses of commonly used phosphatidylcholines as a function of temperature. \textit{Biochimica et Biophysica Acta (BBA) - Biomembranes}. \textbf{1808}, 2761-2771, (2011).
    
    \bibitem{Kucerka2005} Kucerka, N., Tristram-Nagle, S., \& Nagle, J. F. Structure of fully hydrated fluid phase lipid bilayers with monounsaturated chains. \textit{The Journal of Membrane Biology}. \textbf{208}, 193-202, (2006).
    
    \bibitem{Filippov2003} Filippov, A. V., Orädd, G., \& Lindblom, G. The effect of cholesterol on the lateral diffusion of phospholipids in oriented bilayers. \textit{Biophysical Journal} \textbf{84}, 3079-3086, (2003).
    
    \bibitem{Simons1997} Simons, K., \& Ikonen, E. Functional rafts in cell membranes. \textit{Nature}. \textbf{387}, 569-572, (1997).
    
    \bibitem{Carpenter2014} Marrink, S. J., Corradi, V., Souza, P. C. T., Ingólfsson, H. I., Tieleman, D. P. \& Sansom, M. S. P. Computational modeling of realistic cell membranes. \textit{Chem. Rev.} \textbf{119}, 6184–6226, (2019).
    
    \bibitem{Sezgin2017} Sezgin, E., Levental, I., Mayor, S., \& Eggeling, C. The mystery of membrane organization: Composition, regulation and roles of lipid rafts. \textit{Nature Reviews Molecular Cell Biology} \textbf{18}, 361-374, (2017).
    
    \bibitem{Torchilin2005} Torchilin, V. P. Recent advances with liposomes as pharmaceutical carriers. \textit{Nature Reviews Drug Discovery}. \textbf{4}, 145-160, (2005).
    
    \bibitem{Funari2003} Funari, S. S., Barceló, F., \& Escribá, P. V. Effects of oleic acid and its congeners, elaidic and stearic acids, on the structural properties of phosphatidylethanolamine membranes. \textit{Journal of Lipid Research} \textbf{44}, 567-575, (2003).
    
    \bibitem{deJong2013} de Jong, D. H., Singh, G., Bennett, W. F. D., Arnarez, C., Wassenaar, T. A., Schafer, L. V., Periole, X., Tieleman, D. P., \& Marrink, S. J. Improved parameters for the Martini coarse-grained protein force field. \textit{J. Chem. Theory Comput.} \textbf{9}, 687-697, (2013)
    
    \bibitem{VanDerSpoel2005} Van Der Spoel, D., Lindahl, E., Hess, B., Groenhof, G., Mark, A. E., \& Berendsen, H. J. C. GROMACS: Fast, flexible, and free. \textit{Journal of Computational Chemistry}. \textbf{26}, 1701-1718, (2005).
    
    \bibitem{Yesylevskyy2010} Yesylevskyy, S. O., Schäfer, L. V., Sengupta, D., \& Marrink, S. J. Polarizable water model for the coarse-grained MARTINI force field. \textit{PLoS Computational Biology}. \textbf{6}, e1000810, (2010).
    \bibitem{Pall2013} Páll, S., \& Hess, B. (2013). A flexible algorithm for calculating pair interactions on SIMD architectures. \textit{Computer Physics Communications} \textbf{184}, 2641-2650, (2013)
    
    \bibitem{Neale2016} Neale, C., \& Pomès, R. Sampling errors in free energy simulations of small molecules in lipid bilayers. \textit{Biochimica et Biophysica Acta (BBA) - Biomembranes} \textbf{1858}, 2539-2548, (2016).
    \bibitem{Bussi2007} Bussi, G., Donadio, D., \& Parrinello, M. Canonical sampling through velocity rescaling. \textit{The Journal of Chemical Physics}. \textbf{126}, 014101, (2007).
    
    \bibitem{Parrinello1981} Parrinello, M., \& Rahman, A. Polymorphic transitions in single crystals: A new molecular dynamics method. \textit{J. Appl. Phys.} \textbf{52}, 7182-7190, (1981).
    
    \bibitem{Patra2003} Patra, M., Karttunen, M., Hyvönen, M. T., Falck, E., Lindqvist, P., \& Vattulainen, I. Molecular dynamics simulations of lipid bilayers: Major artifacts due to truncating electrostatic interactions. \textit{Biophysical Journal} \textbf{84}, 3636-3645, (2003).
    
    \bibitem{Hess1997} Hess, B., Bekker, H., Berendsen, H. J. C., \& Fraaije, J. G. E. M. LINCS: A linear constraint solver for molecular simulations. \textit{Journal of Computational Chemistry}. \textbf{18}, 1463-1472, (1997).
    
    \bibitem{Lewis1983} Lewis, B. A., \& Engelman, D. M. Lipid bilayer thickness varies linearly with acyl chain length in fluid phosphatidylcholine vesicles. \textit{Journal of Molecular Biology}. \textbf{166}, 211-217, (1983).
    \bibitem{Cullis1979} Cullis, P. R., \& de Kruijff, B. Lipid polymorphism and the functional roles of lipids in biological membranes. \textit{Biochimica et Biophysica Acta (BBA) - Reviews on Biomembranes}. \textbf{559}, 399-420, (1979).
\end{thebibliography}
\end{document}